\documentstyle[twocolumn,aps]{revtex}
\begin{document}
\input epsf.sty
\bibliographystyle{roman}
\def\hb{\hfill\break}
\def\MeV{\rm MeV}
\def\GeV{\rm GeV}
\def\TeV{\rm TeV}

\def\m{\rm m}
\def\cm{\rm cm}
\def\mm{\rm mm}
\def\lam{$\lambda_{\rm int}$}
\def\rad{$X_0$}
 
\def\NIM{Nucl. Instr. and Meth.~}
\def\etal{{\it et al.~}}
\def\eg{{\it e.g.,~}}
\def\ie{{\it i.e.~}}
\def\cf{{\it cf.~}}
\def\etc{{\it etc.~}}
\def\vs{{\it vs.~}}
\begin{minipage} [t]{6.7in}
\Large 
\centerline{\bf On the Accelerated Expansion of the Universe}
\vskip 4mm
\normalsize
\centerline{Richard WIGMANS}
\vskip 2mm
\centerline{\em Department of Physics, Texas Tech University, Lubbock TX 79409-1051, USA}
\vskip 2mm
\centerline{(Submitted to Phys. Rev. Lett. on September 1, 2004)} 
\vskip 3mm

\normalsize

\begin{abstract}
The Universe is filled with relic neutrinos, remnants from the Leptonic Era. Since the formation of galaxies
started, gravitation has modified the Fermi-Dirac momentum distribution of these otherwise decoupled
particles. Decelerated neutrinos moving toward the field-free regions between galaxies could violate the Pauli
principle. The fermion degeneracy pressure resulting from this leads to an accelerated motion of galaxies away
from one another. We show that this model not only offers a natural explanation for the accelerated expansion of
the Universe, but also allows a straightforward calculation of the Hubble constant and the
time-evolution of this constant. Moreover, it sets a lower limit for the (average) neutrino mass. For the latter,
we find
$m_\nu > 0.25$ eV/$c^2$ (95\% C.L).
\hb\hb 
\end{abstract}
\end{minipage}
\vskip 12mm


The experimental observation \cite{Rie98,Per99} that distant Type Ia Supernovae are
dimmer than expected seems to lead to the inevitable conclusion that
the rate at which the Universe expands has increased since the time when these stellar explosions occurred,
5 - 10 billion years ago. 
Current cosmological reviews \cite{PDG04} ascribe the responsibility for this phenomenon to 
anti-gravitational action associated with ``dark energy''. 
However, the nature of this energy and, therefore, the meaning of the non-zero cosmological constant which is
needed in the equations that describe the evolution of the Universe is a mystery.

In this letter, we argue that there is another scenario that may explain the experimental observations. This
scenario does not invoke new forces or unknown forms of energy. It is based on a well known phenomenon, the
{\em fermion degeneracy pressure}, a consequence of Pauli's Exclusion Principle. This degeneracy pressure is
responsible for a variety of astrophysical phenomena, \eg the characteristics of White Dwarfs and neutron stars.
Whereas the fermions involved in these objects are electrons and neutrons, the ones responsible for the degeneracy
pressure discussed here are neutrinos.  
The proposed scenario requires that neutrinos have masses of $\sim 1$ eV/$c^2$.
\vskip 5mm
  
%
According to the Big Bang model, large
numbers of neutrinos and antineutrinos have been around since the earliest stages of the evolving Universe. 
Since the decoupling that marked the end of the Leptonic Era, 
the wavelengths of these relic (anti-)neutrinos have been expanding in
proportion to the size of the Universe. Their present spectrum is believed
to be a momentum-redshifted relativistic Fermi--Dirac distribution, and the 
number of particles per unit
volume in the momentum bin $(p,p+dp)$ is given by
\begin{equation}
N(p) dp~=~{8\pi{p^2 dp}\over{h^3 [\exp (pc/kT_{\nu}) + 1]}}{\bigl({g_\nu\over 2}\bigr)}
\label{numom}
\end{equation}
\vskip 200mm
\centerline{  }
\vskip 67mm\parindent=0mm
\begin{figure}[bthp]
\epsfxsize=87mm \epsfbox{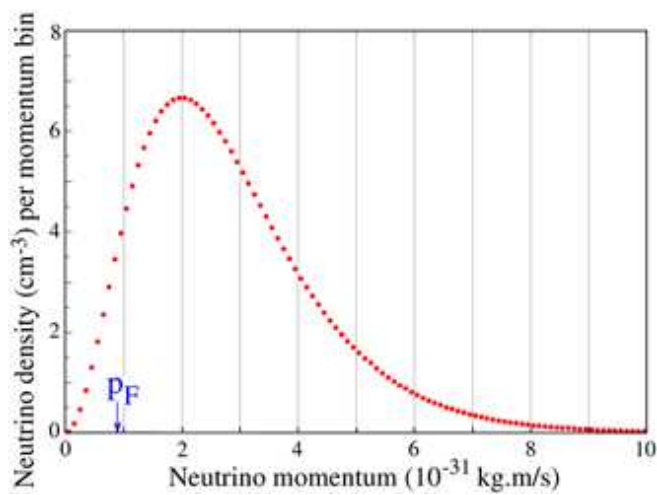}
\vspace{0.5cm}
\caption{\small
The momentum distribution of relic neutrinos at a temperature $T_{\nu} = 1.95$ K. The number of helicity states is
assumed to be 2.}
\label{nuspec}
\end{figure}
where $g_\nu$ denotes the number of neutrino helicity states \cite{TG}.
This momentum spectrum is depicted in Figure \ref{nuspec}.

The distribution is
characterized by a temperature $T_\nu$, which is somewhat lower than that for the relic photons, which were
reheated when the electrons and positrons decoupled. Since 
$(T_\nu/T_\gamma)^3 = 4/11$ and $T_\gamma = 2.725 \pm 0.001$ K \cite{COBE}, $T_{\nu}$ is expected to be 1.95 K.
For a neutrino temperature of 1.95 K, the Fermi momentum ($p_F = kT_\nu/c$) is $1.68 \cdot 10^{-4}$ eV/$c$, or 
$9.0\cdot 10^{-32}$ kg.m.s$^{-1}$.

The present density of these Big Bang relics is estimated at $\sim$ 220 cm$^{-3}$, for each (Dirac) neutrino
flavor \cite{boehm}, nine orders of magnitude larger than the density of baryons in the Universe. 

It is important to realize that, depending on their mass, these relic neutrinos might
be very {\em nonrelativistic} at the current temperature. Since they decoupled,
their momenta have been stretched by a factor of $10^{10}$, from\break 
1 MeV/$c$ to $10^{-4}$ eV/$c$. If their rest
mass were 1 eV/$c^2$, then their Fermi velocity ($v_F = p_F/m_\nu$) would thus be $1.68 \cdot 10^{-4} c$, or only
50 km/s.

The experimental upper limit on the mass of the electron antineutrino was recently determined at 2.2
eV/$c^2$ (95\% C.L.), from a study of the electron spectrum of $^3$H decay \cite{Mainz}. The
experimental results on atmospheric and solar neutrinos obtained by the Superkamiokande \cite{SuperK}
and SNO\cite{SNO} Collaborations suggest that neutrinos do have a non-zero rest mass. A conservative
interpretation of the experimental results is that at least one of the neutrino mass eigenvalues is larger
than 0.04 eV/$c^2$. There is no experimental information that rules out a neutrino rest mass of the order of  0.1
-- 1 eV/$c^2$.
\vskip 5mm


During the radiation era, the neutrino spectrum was only affected by the gradual expansion, with $p$ and $T_{\nu}$
inversely proportional to the evolving distance scale. 
However, when gravity became the dominant force in the Universe and stars and galaxies started to
form, important changes were about to take place.

Since neutrinos are subject to gravitational forces, their spectra were affected. The neutrino velocities either
increased or decreased as a result of gravitational acceleration or deceleration, depending on the direction of
motion of the particles with respect to the (dominant) gravitational source. The effects of this extended over
intergalactic distances, to regions far away from these sources. For example, a neutrino moving away from our
galaxy ($M \approx 10^{11} M_\odot$) at a distance of 300 kpc ($10^{22}$m), experiences a gravitational
deceleration
of $\sim 10^{-13}$ m.s$^{-2}$. If the initial velocity of this neutrino was 50 km/s, it would 
over a period of 5 billion years (the relevant time scale for the issues discussed in
this letter) be slowed down by about a factor of 2 as a result of this deceleration.
The velocity distribution of the relic neutrinos
has thus gradually changed in the non-uniform gravitational fields that resulted from baryon clustering, \ie
galaxy formation. 

\begin{figure}[bthp]
\hspace{0.5cm}
\epsfxsize=67mm \epsfbox{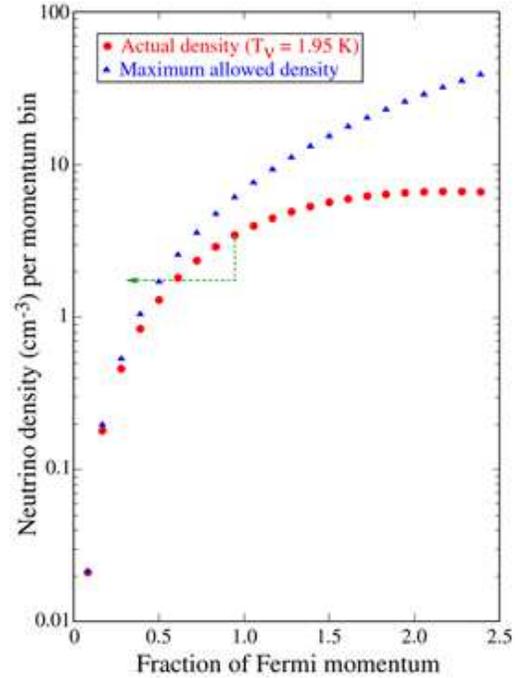}
\vspace{0.5cm}
\caption{\small
The actual number of quantum states per cm$^3$ occupied by relic neutrinos 
at $T_\nu = 1.95$ K, and the maximum available density, as a function of the neutrino momentum (in bins
of 0.11 $p_F$). The arrow indicates what happens when half of the neutrinos in a certain momentum bin are
decelerated.}
\label{nufree}
\end{figure}

To understand the potential problems that may be caused by this, it is important to realize that the relic
neutrinos form a degenerate fermion gas, at $T_\nu = 1.95$ K. Especially at momenta $p \ll p_F$, almost all
available quantum states are occupied (Figure \ref{nufree}). 
The expansion of the Universe does not change that fact, since the neutrino momenta are inversely proportional
to the Universal scale and the maximum allowed fermion density evolves in proportion to $p^3$. The present
situation is thus not new. The Universe has {\em always} been filled with a degenerate neutrino gas.

The neutrinos that have lost the largest fraction of their momentum by gravitational deceleration are those 
whose velocities were close to the escape velocity when galaxy formation started.
They tend to concentrate in the low-field regions surrounding the center-of-mass of galaxy clusters.
If they have been sufficiently decelerated, then there may not be enough quantum
states there to accommodate them. A constant or decelerated Hubble expansion does not prevent, alleviate or solve this
problem.

So what happens when gravitationally decelerated neutrinos cannot find a quantum state to fit into? 
Any attempt to squeeze more fermions into the available volume than allowed by the limits set by the Pauli
principle creates a pressure that prevents this from happening. 
In the case of White Dwarfs and neutron stars, this pressure prevents the gravitational collapse of these objects.
{\em In a cluster of galaxies, this fermion degeneracy pressure leads to an accelerated motion of the galaxies
away from their common center-of-mass}.  

Let us consider two galaxies of equal mass $M$, separated by a distance $2r$. 
The degeneracy pressure prevents decelerated neutrinos from violating the Pauli principle in the low-field
region surrounding the center-of-mass point ($C$) halfway between these galaxies.
This is accomplished by curving the space around $C$ such that the inertial reference frames of the galaxies
responsible for the deceleration experience a compensating force in the direction away from $C$.  
The accelerated motion resulting from this is the {\em only way} to achieve that none of the
${\cal{O}}(10^{77})$ neutrinos populating the cubic Mpc surrounding a typical galaxy makes a ``soft landing'' in
the region around $C$. In this way, gravitationally decelerated neutrinos which otherwise would
range out in this ``forbidden'' region will always feel a force pushing them away from it.  
 
The
acceleration is thus equal to the deceleration a particle in $C$ would feel in the absence of other galaxies.    
Since this acceleration, which we will call the Pauli acceleration in the following, takes place with respect to
the center-of-mass ($a = GMr^{-2}$), it is four times larger than the gravitational deceleration of the two
galaxies with respect to each other. Since they are separated by a distance $2r$, this deceleration is
$a = -GM(2r)^{-2}$. The net result is thus that the galaxies move
apart at an {\em increasing} speed, regardless of their mass or distance.
\vskip 5mm


Because it is necessitated by gravitationally decelerated relic neutrinos, the Pauli acceleration is a relatively
recent phenomenon. It only started to play a role after the first galaxies had formed and the resulting non-uniform
gravitational fields had sufficiently reduced the speed of some fraction of the relic neutrinos.

Based on the scenario described above, we have simulated the expansion, tracing it back to its
beginnings. We have taken two galaxies separated by a distance ($R_0$) of 1 Mpc, typical for the present Universe.
These galaxies move apart at a relative velocity of 73 km/s, \ie the current value of the Hubble constant,
$H_0$ \cite{PDG04}. We follow the history of this system back in time, in small steps
(0.1\% of the age of the Universe at each point). As we go back in time, the relative velocity decreases as a
result of the Pauli acceleration, and as the galaxies get closer, this acceleration increases.
\begin{figure}[bthp]
\epsfxsize=87mm \epsfbox{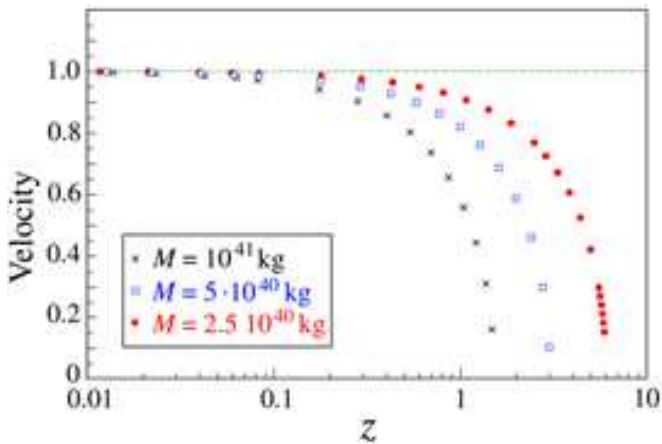}
\vspace{0.5cm}
\caption{\small
The relative velocity at which two galaxies currently separated by 1 Mpc are moving apart, as a function of the
redshift, for different values of the galaxy mass. Only the Pauli acceleration has been taken into account.}
\label{zmspeed}
\end{figure}

Figure \ref{zmspeed} shows some results of this study. The relative velocity of the two galaxies is shown as a
function of the redshift $z$, which in this context is simply the ratio $R_0/R(-t)$. The results turned out to
be quite sensitive to the chosen value of the galaxy mass. A change of a factor of 2 in this mass made a
considerable difference for the $z$ value at which the expansion started to play a role. In the next round of
simulations, we chose this mass according to the best current estimates \cite{PDG04}, which have the total matter
density at $\sim 25\%$ of the critical value ($\Omega_m = 0.25$). Therefore,
$$\rho_m = {{3\Omega_m H^2}\over {8\pi G}} = 2.5\cdot 10^{-27}~~{\rm kg.m}^{-3}.$$
This gives for the total mass contained in a sphere with a diameter of 1 Mpc $\sim 4\cdot 10^{40}$ kg.    
\begin{figure}[bthp]
\epsfxsize=87mm \epsfbox{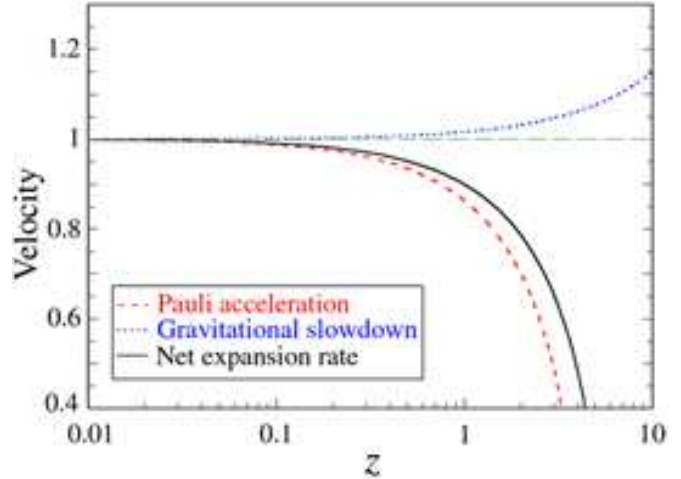}
\vspace{0.5cm}
\caption{\small
The relative velocity at which two galaxies currently separated by 1 Mpc are moving apart, as a function of the
redshift. The mass of each galaxy is $4\cdot 10^{40}$ kg.}
\label{expan}
\end{figure}

Figure \ref{expan} shows the results of our simulations for this choice of galaxy masses. The dashed and dotted
lines describe the separate effects of the Pauli acceleration and the gravitational slowdown, the solid curve
represents the combined effect. These results, which imply a net
increase of the expansion velocity by $\sim 10\%$ in the period since $z = 1$, are in agreement with the
Supernova data.

These results also suggest that the Pauli acceleration started around $z = 6$, which is not unreasonable given
the current ideas about the start of galaxy formation \cite{PDG04}. We can also follow the development in
the reverse order and thus calculate the current expansion rate. 
Figure \ref{Hdnow} shows this expansion rate, \ie the current value of $H_0$, as a function of the redshift
at which the Pauli acceleration started.  
\vskip 5mm

The described expansion scenario has several interesting consequences for the present structure of the Universe.
We mention two that seem to be confirmed by experimental observations:
\begin{enumerate}
\item The distribution of galaxies in the Universe is non-uniform. Field-free regions have the tendency to grow
over time, since they push all galaxies in their vicinity away with increasing velocities. One should therefore
expect large voids, surrounded by strings of galaxies.
\item Each galaxy is surrounded by a ``halo'' of gravitationally trapped neutrinos. This halo may extend\break
\begin{figure}[bthp]
\epsfxsize=87mm \epsfbox{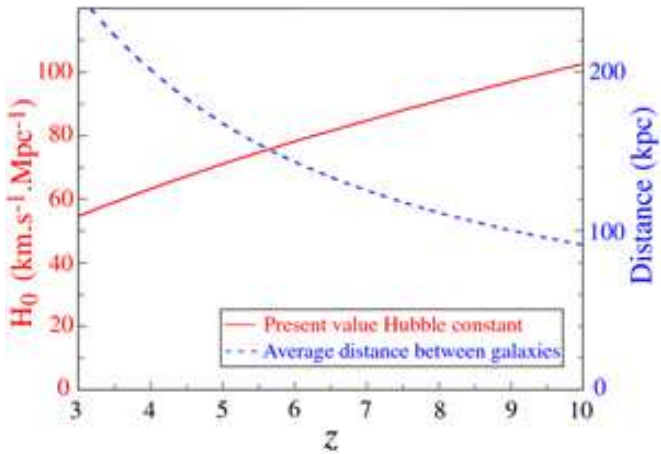}
\vspace{0.5cm}
\caption{\small
The relative velocity at which two galaxies, currently separated by 1 Mpc, are moving apart ({\em left hand
scale}), and the distance separating the galaxies ({\em right hand scale}) as a function of the redshift at which
the Pauli acceleration started. The mass of each galaxy is
$4\cdot 10^{40}$ kg.}
\label{Hdnow}
\end{figure}
over a distance of up to several hundred kpc. The density of available quantum states is proportional to $r^{-3/2}$
($v_{\rm esc}^3$) and, therefore, the total mass in the galaxy plane may be expected to scale
as $1/\sqrt{r}$. This halo may represent a substantial fraction of the dark matter. How large this fraction is depends
on the neutrino rest mass.
\end{enumerate}
\vskip 5mm

The scenario described in this letter hinges of course critically on the value of the neutrino
mass. If this mass were too small, then the gravitational effects on the relic neutrino velocity would be
insignificant. We have estimated the lower limit on $m_\nu$ as follows.

In order to violate the Pauli principle, relic neutrinos would have to lose at least one third of their
momentum through gravitational deceleration. This is illustrated by the arrow in Figure \ref{nufree}, which
indicates what happens when half of the neutrinos in a certain momentum bin are decelerated (the
other half are accelerated). Neutrinos with momenta larger than $\sim 1.5 p_F$ even have to lose more than half
of their momentum. 

According to the neutrino
spectrum described by Equation \ref{numom} and shown in Figure
\ref{nuspec}, 95\% of the relic neutrinos have velocities larger than $0.82 v_F$. These neutrinos have to lose at
least 40\% of their momentum before they could trigger the fermion degeneracy pressure. 

Simulations of the type described above showed that when two galaxies with a standard mass of 
$4\cdot 10^{40}$ kg each move apart at a rate of 73 km/s, neutrinos with velocities 
up to 165 km/s (measured in the rest frame of one of the galaxies) may lose 40\% or more of their momentum
through gravitational deceleration. Therefore, the Fermi velocity of the neutrinos that may accomplish this
feat must be smaller than 200 km/s. And since 
$m_\nu$ (in units of eV/$c^2$) equals $1.68\cdot 10^{-4} c/v_F$, this means that
the (average) neutrino mass has to be larger than 0.25 eV/$c^2$ (95\% confidence level).   
\vskip 5mm

Fermion degeneracy pressure is a phenomenon that comes into action whenever the fermion density approaches the
limits set by the Pauli Exclusion Principle. Until now, it has been exclusively associated with the extremely high
temperatures and densities that characterize the interior of degenerate stellar objects. It is remarkable that the
same phenomenon may also play a crucial role in the extremely cold and empty conditions of intergalactic space.

We have shown that neutrino degeneracy pressure may lead to an accelerated expansion of the Universe, which
would explain not only the Supernovae Type Ia data but also the current value of $H_0$. If this explanation is
correct, then the expansion of the Universe will continue forever, since the driving principle is not going to
go away. It also means that planned efforts to improve the neutrino mass sensitivity in studies of $^3$H decay
to the 0.2 eV/$c^2$ level \cite{Mainz} may well pay off in the form of a precise measurement of this important
parameter. 

\section*{Acknowledgment}

The author would like to thank the Istituto Nazionale di Fisica Nucleare in Frascati, Italy, and in particular its
director, Dr. Sergio Bertolucci, for their hospitality during a sabbatical stay and for the opportunity to think
about the Universe in an inspiring environment.
 
\bibliographystyle{unsrt}

\end{document}